\documentclass[11pt]{article}
\usepackage{amssymb}
\usepackage{epsfig}
\newlength{\bredde}
\def\slash#1{\settowidth{\bredde}{$#1$}\ifmmode\,\raisebox{.15ex}{/}
\hspace*{-\bredde} #1\else$\,\raisebox{.15ex}{/}\hspace*{-\bredde} #1$\fi}
\newcommand{\beq}{\begin{equation}}
\newcommand{\eeq}{\end{equation}}
\newcommand{\beqn}{\begin{eqnarray}}
\newcommand{\eeqn}{\end{eqnarray}}

\newcommand{\al}{\alpha}
\newcommand{\la}{\lambda}

\newcommand{\ka}{\kappa}
\newcommand{\ga}{\gamma}
\newcommand{\si}{\sigma}
\newcommand{\dga}{\frac{\partial}{\partial g}}
\newcommand{\sect}[1]{\setcounter{equation}{0}\section{#1}}

\newcommand{\Pint}{{\mathbf{-}}\!\!\!\!\!\!\int}

\begin{document}
\topmargin -1.4cm
\oddsidemargin -0.8cm
\evensidemargin -0.8cm
\title{\Large{{\bf 
New Critical Matrix Models and Generalized Universality}}}

\vspace{1.5cm}

\author{~\\{\sc G. Akemann}\\~\\
CEA/Saclay, Service de Physique 
Th\'eorique\footnote{Unit\'e associ\'ee CNRS/SPM/URA 2306}\\
F-91191 Gif-sur-Yvette Cedex, France
\\{\it e-mail: akemann@spht.saclay.cea.fr}
\\~\\and\\~\\
{\sc G. Vernizzi}\\~\\
Department of Theoretical Physics, Oxford University\\
1 Keble Road, Oxford, OX1 3NP, United Kingdom\\
{\it e-mail: vernizzi@thphys.ox.ac.uk}}
\date{}
\maketitle
\vfill
\begin{abstract}
We study a class of one-matrix models with an action containing
nonpolynomial terms. By tuning the coupling constants in the action to
criticality we obtain that the eigenvalue density vanishes as an
arbitrary real power at the origin, thus defining a new class of
multicritical matrix models. The corresponding microscopic scaling law
is given and possible applications to the chiral phase transition in
QCD are discussed. For generic coupling constants off-criticality we
prove that all microscopic correlation functions at the origin of the
spectrum remain in the known Bessel universality class.  An arbitrary
number of Dirac mass terms can be included and the corresponding
massive universality is maintained as well. We also investigate the
critical behavior at the edge of the spectrum: there, in contrast to
the behavior at the origin, we find the same critical exponents as
derived from matrix models with a polynomial action.
\end{abstract}
\vfill

\begin{flushleft}
SPhT T01/141\\
OUTP-01-62P\\
hep-th/0201165
\end{flushleft}
\thispagestyle{empty}
\newpage

\renewcommand{\thefootnote}{\arabic{footnote}}
\setcounter{footnote}{0}

\sect{Introduction}\label{intro}

The study of random matrix models as simplified versions of more
complicated interacting Quantum field theories has often been very
fruitful. For instance, there are zero-dimensional matrix models with
applications ranging from Quantum Gravity to Quantum Chromodynamics
(QCD).  For a review on these topics we refer to \cite{Philippe} and
\cite{JT}, respectively, as well as to \cite{GMW} for applications in
other fields.  In most of the cases it has been sufficient to consider
the so-called Gaussian matrix model, where the random matrices (that
describe for example an underlying Hamiltonian) have entries which are
Gaussian random variables.  The reason for such Gaussian matrix models
being generic is their property of universality. Universality in
random matrix models means that when reweighting the matrix elements
differently within a certain class of perturbations around the
Gaussian models the same correlation functions of the matrix
eigenvalues are retained. Different proofs of universality
\cite{AJM,BZ,ADMN,KF97b,KF97,DN,DNII,SV,AK} using various techniques
have been obtained in distinct large-$N$ regimes, where $N$ is the
size of the matrices.

In general one has to distinguish between the {\it macroscopic}
large-$N$ limit where the oscillations of the correlation functions
are smoothed and the {\it microscopic} large-$N$ limit, where the
eigenvalues are rescaled such that a particular region of the spectral
density is magnified. This region might be the origin, the bulk or the
endpoint of the spectrum and microscopic universality classes emerge
correspondingly. In fact the universal kernel of orthogonal
polynomials that generates all the correlation functions is described
by Bessel, Sine or Airy functions, respectively.  The stability of
universality under changes of the functional form of the measure has
also been tested \cite{kim,us} and again the microscopic and
macroscopic regime behave differently. In this paper we will see
another extension of universality by adding nonpolynomial terms to
Gaussian (or more general) weight functions.

However, this is not the main motivation of the present work. In many
applications one has to take into account extensions of the Gaussian
model and to fine tune in order to reach critical points (for a review
on phase transitions in matrix models we refer to
\cite{GCicuta}). There may be several different reasons for doing
so. For instance, in Quantum Gravity the critical behavior is used to
enhance the otherwise subleading contributions in the $1/N^2$
topological expansion and make them all contribute in the double
scaling limit (for a review see \cite{Philippe}).  In the study of
chiral symmetry breaking in QCD
\cite{SV93,JT} the spectral density at the 
origin $\rho(0)$ is proportional to the order parameter, the chiral
condensate. Terms of higher order than Gaussian may then be used to
achieve a phase transition by requiring $\rho(0)=0$ \cite{ADMNII}.
The corresponding scaling exponents and correlation functions will be
those on top of the phase transition. Such matrix models have already
been investigated in \cite{ADMNII} where also universality was found
within a given multicriticality class.  But only very specific
transition points could be reached in \cite{ADMNII}, where
$\rho(\la=0)\sim\la^{2m}$ with $m\in\mathbb{N}$. Our aim here is to
complete this study and find new critical models where the density
vanishes with an arbitrary real power. For that purpose we will add a
nonpolynomial part to a Gaussian potential, or to a more general
polynomial potential, and study its effect.  It has been argued in
\cite{JV} that at the temperature induced chiral phase transition in
QCD the eigenvalue density precisely behaves as
$\rho(\la=0)\sim\la^{1/\delta}$, with mean field exponent $\delta=3$.

Single nonpolynomial potentials have been already studied in
\cite{Pastur,CWK,FKY,Ken}.  They have been shown to fall into the microscopic
universality class of the sine kernel in the bulk and at the origin of
the spectrum.  It does therefore not come as a surprise that the
extensions of the model we study here fall into the corresponding
massless and massive Bessel universality classes \cite{ADMN,DN,DNII}
when Dirac mass terms are included. This holds as long as we keep away
from criticality or in other words when we maintain $\rho(0)\neq 0$.
The investigation of new multicritical points is however only possible
with several interaction terms in the potential, polynomial and
nonpolynomial.  This is our starting point.

Nonpolynomial potentials have also been studied in the context of 
Quantum Gravity \cite{GM,IK}.
We will analyze whether the terms we introduce may
also change the critical exponents \cite{Kazakov} in the 
appropriate large-$N$ limit.
Finding real critical exponents would correspond to new
representations of conformal field theories by simple one-matrix
models. However, our findings in this respect are negative in the
sense that the perturbations we consider here will not lead out of the
known universality classes. Nevertheless it would be very interesting
to find realizations of such simple one-matrix models instead of more
complicated multi-matrix or $O(n)$ models
\cite{Daul,Eynard}.

The matrix-model approach to the chiral phase transition we described
so far is not the only possible one. The phase transition can also be
studied by adding an external field \cite{JV,BH98} or by allowing more
substructure for the matrices such as additional spin or gauge group
degrees of freedom \cite{Benoit}. While such extensions are more close
to the phenomenology they are certainly more difficult to solve
exactly without any further assumptions. However, let us emphasize
that such a matrix model \cite{HJSSV} was used to deduce the phase
diagram of QCD realistic for two massless quarks in the
temperature-density plane and that other approaches have confirmed
this general picture (see e.g. \cite{SRS}).  It consists of a
second-order line beginning at a critical temperature with zero density
that joins a first-order line starting from a critical density at
zero temperature in a tricritical point.

In the multicritical matrix models considered in \cite{ADMNII} the
phase transition is of third order.  Therefore these models may become
interesting precisely at the tricritical point. They may also be
relevant for the chiral phase transition occurring when the number of
flavors $N_f$ is raised above the critical point, where the
beta-function of QCD changes its sign. The relation of the large-$N_f$
phase transition to the one occurring in the Gross-Witten model has
been pointed out in \cite{poulGW}. 
The extension of \cite{ADMNII} to nonpolynomial potentials we
investigate here may also have a phase transition of different order.
The fact that the critical exponents of \cite{JV} are included in our
model suggests a possible second order transition.
However, the continuity of our results at even integer powers
suggests that the transition remains of third order.
We have not been able to give a definite answer to that question.

The paper is organized as follows.  We will mainly concentrate on
Hermitian matrix models also called the Unitary Ensemble. This model
can be used to study flavor symmetry breaking in three dimensional QCD
\cite{VZ94}. In the original approach to chiral symmetry breaking in
four-dimensional QCD \cite{SV93} a slightly different matrix model is
needed, the chiral Unitary Ensemble.  In order to translate our
results to the chiral ensemble we will simply use the relations
derived in \cite{ADMNII} and \cite{ADIII} between the corresponding
orthogonal polynomials and kernels.  In Section
\ref{largeN} we first state our model with arbitrary polynomial and
nonpolynomial power like terms in the potential. It includes Dirac
mass terms motivated by the application to QCD.  These terms are also
studied under the name of characteristic polynomials.  In the same
section we calculate the macroscopic large-$N$ spectral density away
from criticality for potentials containing nonpolynomial parts using
different techniques.  We also prepare the ground for Section
\ref{univ} by recalling the orthogonal polynomial method in the
version as being reviewed in \cite{KFrev}.  In Section \ref{univ} we
present our results about generalized universality. We prove that away
from criticality the correlation functions in the 
microscopic scaling limit at the origin  
fall into the same massless and massive universality classes as in
\cite{ADMN,DN,DNII}, also in the presence of nonpolynomial terms. The
tool \cite{KFrev} we use is to derive a universal differential
equation for the orthogonal polynomials first without massive flavors.
Then, in order to include an arbitrary number of Dirac masses we
proceed along the lines of \cite{AK} without any restriction on the
number of flavors.  Our proof is valid for ensembles with chiral and
nonchiral unitary invariance and is an alternative to the original
proofs \cite{DN,DNII} for the massive ensembles.  In Section
\ref{multi} we come to our main result by tuning the coupling
constants of our model to criticality. After describing in detail the
two simplest examples we introduce a general class of new
multicritical densities. They behave as $\rho(\la=0)\sim\la^{\ka-1}$
with real noninteger $\ka>1$.
For $2m-1<\ka<2m+1$, $m\in\mathbb{N}$, we also give the corresponding minimal 
critical potentials containing $m$ polynomial terms and one nonpolynomial 
term.
In Section \ref{softmulti} we investigate possible critical points in
the leading order of the free energy at large-$N$ for the simplest of
such multicritical models. We find a nonanalytic behavior at large-$N$
when the spectral density is tuned to develop additional zeros at the
edge of the spectrum. The critical exponent $\gamma_{str}=-\frac12$ we
find for the free energy is the same as known from matrix models with
polynomial potentials
\cite{Kazakov}.
Section \ref{con} contains our conclusions and in the Appendix we
collect some results on the so-called two-cut spectral density, where
the eigenvalues lie on two disjoint intervals that merge at the
transition.


\sect{The model and its large-$N$ solution}\label{largeN}

We start by describing the most general matrix model we wish to work
with. In order to make sense of nonpolynomial terms we give the model
right away in terms of the $N$ eigenvalues of a Hermitian matrix:
\beqn
{\cal Z}^{(2N_f,\al)}(\{m_f\}) 
&=& \int_{-\infty}^\infty \prod_{i=1}^N 
\left( d\la_i\  |\la_i|^{2\al}
\prod_{f=1}^{N_f}\left(\la_i^2\ +\ m_f^2\right)\, 
e^{-N V(\la_i) }\right)
|\Delta(\la)|^2 \ ,  
\label{Z}\\
V(\la) &\equiv& V_{pol}(\la) + V_{nonpol}(\la) \ \equiv\ 
\sum_{j=1}^{d_p} \frac{g_{2j}}{2j} \la^{2j} \ +\ 
         \sum_{j=1}^{d_{np}} {h_j} |\la|^{\ka_j} \ \ \ ,
\label{pot}
\eeqn
where $1<\ka_1<\ldots<\ka_{d_{np}}\in\mathbb{R}\backslash\mathbb{N}$.
Here we have restricted ourselves to $1<\ka_j$ as we do not only need
a normalizable spectral density $\rho(\la)$ enforcing $0<\ka_j$, but
also $\rho(0)<\infty$ which requires $1<\ka_j$, as we will see.  The
Vandermonde determinant $\Delta(\la)=\prod_{k>l}^N(\la_k-\la_l)$
originates from the diagonalization of the Hermitian matrix.  The
potential $V(\la)$ is chosen to be symmetric for simplicity and in
view of applications to QCD (as we will show, odd powers are not
necessary for obtaining multicritical potentials).  It has been split
into a polynomial part, $V_{pol}(\la)$, and a nonpolynomial part,
$V_{nonpol}(\la)$. In the latter we have allowed for an arbitrary
number $d_{np}$ of real positive powers $\ka_j$.  In order to
achieve a multicritical behavior at the origin we will actually need
only one of such terms. However, for the large-$N$ solution presented
in this section as well as for the issue of universality in the next
section any number of such terms may be present.

The Dirac mass terms in front of the exponential are added in order to
study flavor symmetry breaking in three-dimensional QCD (see
\cite{VZ94} for details). The $2N_f$ Dirac masses occur in $N_f$ pairs
$\pm im_f$. Furthermore, we have included a massless Dirac determinant
raised to some arbitrary real power $2\al>-1$. We will need this term
later in Section
\ref{univ} 
when we switch from the Unitary Ensemble of Hermitian matrices in eq. 
(\ref{Z}) to the chiral Unitary Ensemble which is needed in applications 
to four-dimensional QCD.

In the following we will determine the spectral density $\rho(\la)$ in
the macroscopic large-$N$ limit. A first argument how to calculate it
in the presence of nonpolynomial terms comes from a saddle-point
analysis. To begin with we observe that the preexponential factors
containing the mass terms in eq. (\ref{Z}) can be formally included
into the potential by shifting $V(\la) \to V(\la) - 1/N
\ln[|\la|^{2\al}
\prod_{f=1}^{N_f}(\la^2+m_f^2)]$. Hence they are subleading in 
the large-$N$ limit and we can set $\al=N_f=0$ in the rest of this section. 
Only in the microscopic limit they will become important while the 
macroscopic spectral density does not depend on  
$\al,N_f\neq 0$.

The spectral density for a single nonpolynomial potential
$V(\la)=|\la|^\ka$ has been calculated already in
\cite{Pastur,CWK,FKY,Ken}. In order to determine the full density $\rho(\la)$
for the potential eq. (\ref{pot}) we make use of the fact that the
saddle-point equation for the spectral density is linear in the
potential. We can therefore obtain solutions of the general problem by
simply adding up solutions for the pure polynomial case with solutions
for the nonpolynomial case.  In order to see that in more details we
repeat some well known facts about the saddle-point solution of matrix
models (for a review see e.g.
\cite{Philippe}). 
Defining the free energy as ${\cal F} \equiv -\ln[{\cal Z}]/ N^2$ and
switching from discrete eigenvalues to a continuous density one has
\beq
{\cal F}\ =\ \int_\si d\la \rho(\la) V(\la) -\int_\si d\la\, d\mu
\rho(\la)\rho(\mu)\ln|\la-\mu| + K \left(\int_\si d\la \rho(\la)-a\right) \ .
\label{F}
\eeq
We have introduced a Lagrange multiplier $K$ to enforce the
normalization of the spectral density $\rho(\la)$ to $a$, which may
differ from unity. Furthermore, we consider here the case where the
support $\si$ of $\rho(\la)$ is a compact interval, the so-called
one-cut solution.  
For the two-cut solution we refer to the Appendix 
\ref{App}. 
The saddle-point equation then reads
\beq
0\ = \  V(\la) -2 \int_\si d\mu
\rho(\mu)\ln|\la-\mu| + K \ ,\ \ \ \ \la\ \in \si \equiv [-c,c]\ ,
\label{SP}
\eeq
which is often written also after taking the derivative w.r.t. $\la$ as
\beq
\Pint_\si d\mu \frac{ \rho(\mu)}{\la-\mu}= \  \frac{V'(\la)}{2} \, , \quad 
\la \in \si \, ,
\label{SPD}
\eeq
and $\Pint$ is the Principal Value of the integral. The Lagrange
multiplier $K$ can be determined by choosing a particular value of
$\la$ in eq. (\ref{SP}), e.g. $\la=0 \in \si$.  Because of the
linearity of eq. (\ref{SP}) two positive solutions $\rho_1$ and
$\rho_2$ for two given potentials $V_1$ and $V_2$, respectively, can
be simply added under the assumption that they have the same support
$\si$. This gives a new solution $\rho=\rho_1+\rho_2$ for the
potential $V=V_1+V_2$ which is nonnegative on the same support $\si$. 
Thus we can take known solutions of the saddle-point equation for
polynomial and nonpolynomial potentials, fix their support to be equal
and add them. The normalization to unity of the new spectral density,
$\rho(\la)=\rho_{pol}(\la)+\rho_{nonpol}(\la)$, has to be imposed
eventually. This condition uniquely determines the dependence of the
end-point $c$ on the coupling constants in the potential or, which is
the same, it gives a relation among the coupling constants for any
given fixed positive number $c$.\\ Let us give an example. Choosing
the simplest polynomial potential, i.e. a Gaussian potential
$V_{pol}(\la)=\frac12 g_2 \la^2$, the corresponding spectral density
is the ``Wigner semi-circle''
\beq
\rho_{Gauss}(\la) \ =\ \frac{g_2}{2\pi} \sqrt{c^2-\la^2} 
\ ,\ \ \ \ \la\ \in [-c,c]\ .
\label{rhogauss}
\eeq
If one fixes the normalization condition then $g_2$ and $c$ must be
related to each other.  In particular, the normalization condition of
the density to unity reads $\frac14 g_2c^2=1$. Here we consider $c$ as
an arbitrary given positive parameter.
For the nonpolynomial part we choose $V_{nonpol}(\la)=g |\la|^\ka$,
which is called Freud weight in the mathematical literature. The
corresponding spectral density normalized to unity has been calculated
in \cite{Pastur,CWK,Ken} (see also references therein). The un-normalized
density sharing the same one-cut support $[-c,c]$ will be re-obtained
in this section and can be written as (see also \cite{FKY}):
\beqn
\rho_{Freud}(\la) &=& \frac{g \ka H c^{\ka-1}}{\pi}
\int_{\frac{|\la|}{c}}^1 ds \frac{s^{\ka-1}}{\sqrt{s^2-\frac{\la^2}{c^2}}} 
\label{rhofreud} \\
&=& \frac{g \ka H c^{\ka-1} }{\pi} F \left( 1-\frac{\ka}{2},1;
\frac{3}{2};1-\frac{\la^2}{c^2} \right) \sqrt{1-\frac{\la^2}{c^2}}   \, ,
\ \ \ \ \la\ \in [-c,c]\ ,
\label{rhofreudF2}
\eeqn
where
\beq
H \equiv \ 
\frac{\Gamma(\frac{\ka}{2}+\frac12)}{\Gamma(\frac12)\Gamma(\frac{\ka}{2})}
\ =\ \frac{1}{B\left(\frac{\ka}{2},\frac12\right)}
\ , \label{H}
\eeq
$B$ is the Beta function and $F$ is the Hypergeometric function. 
It can be easily seen from the integral
that for $0<\ka$ the density is normalizable and that for 
$1<\ka$ it holds $\rho(0)<\infty$. 
For a discussion of $0<\ka\leq1$ we refer to \cite{CWK,FKY}.
Using some 
identities among Hypergeometric functions for $\ka \neq 1,3,5,\ldots$, eq. 
(\ref{rhofreudF2}) 
can also be written in the useful form:
\beq
\rho_{Freud}(\la)\ =\ 
\frac{g \ka H c^{\ka-1}}{\pi}\left(
\frac{1}{2H}\tan\left[\frac{\ka\pi}{2}\right] 
\left(\frac{\la}{c}\right)^{\ka-1}
+\frac{1}{\ka-1}F\left(\frac12,\,
\frac12-\frac{\ka}{2};\,\frac32-\frac{\ka}{2};\,\frac{\la^2}{c^2}\right)
\right) \, .
\label{rhofreudF}
\eeq
Imposing the condition $1=Hg c^\ka$ we would obtain a normalized density, 
reproducing \cite{Pastur,CWK,Ken}. 
By adding the two solutions eqs. (\ref{rhogauss}) and (\ref{rhofreud}) 
which now have the same support, we obtain for the full spectral density 
\beqn
\rho(\la) &=& \rho_{Gauss}(\la)\ +\ \rho_{Freud}(\la)   \nonumber \\
&=& \frac{g_2}{2\pi}  \sqrt{c^2-\la^2}\ +\ 
\frac{g \ka H c^{\ka-1}}{\pi}
\int_{\frac{|\la|}{c}}^1 ds \frac{s^{\ka-1}}{\sqrt{s^2-\frac{\la^2}{c^2}}} 
\ ,\ \ \ \ \la\ \in [-c,c]\ .
\label{rhoGF}
\eeqn
By imposing $\int_\si d\la \rho(\la)=1$ we obtain the following relation 
between the coupling constants and the endpoint 
$c$ of the support:
\beq
1 \ = \  \frac14 g_2 c^2 + Hg c^\ka \ .
\label{BC}
\eeq 
This relation can be easily checked by comparing to the known polynomial 
potential with quadratic and quartic coupling  \cite{BPIZ} by setting $\ka=4$.
We can immediately draw two consequences. First, both parts of the density, 
$\rho_{Gauss}(\la)$ and $\rho_{Freud}(\la)$ have an expansion in powers of 
$\la^2$ around the 
origin with the latter having an extra term $\sim\la^{\ka-1}$ 
as one can see from eq. (\ref{rhofreudF}). 
Consequently a 
critical behavior at the origin can be achieved by tuning the coupling 
constants $g_2$ and $g$ such that the leading constant term, $\rho(0)$,
vanishes. 
This will be analyzed in great detail in Section \ref{multi}.
Second, it is easy to see how to generalize the above adding of spectral 
densities.  
Since the density for a polynomial potential is already know to 
most generality \cite{AJM}
we can use the above procedure to add successively single
nonpolynomial potentials of higher and higher real power to obtain the density
for the potential eq. (\ref{pot}). \\
A standard way to obtain eqs. (\ref{rhofreud}) and  (\ref{rhofreudF2}), 
is by introducing the resolvent $G(z)$:
\beq
G(z) \equiv \int_{\si} d\mu \frac{\rho(\mu)}{z-\mu} \, , \quad z \in 
\mathbb{C}  \backslash \si
\eeq
in terms of which the saddle point equation (\ref{SPD}) reads 
$G(\la+i 0)+G(\la-i 0)=V'(\la)$, $\la \in \si$. Clearly, $G(z)$ has a cut at 
$z\in\si$.
The (unique) solution which is bounded at the end points and which 
behaves like $G(z) \sim 1/z$ at large $|z|$ is
 (see e.g. \cite{Muskhe})
\beqn
G(z)&=&\sqrt{z^2-c^2} \int_{-c}^c \frac{dt}{2 \pi} \frac{V'(t)}{(z-t) 
\sqrt{c^2-t^2}} \, \nonumber \\
&=& g H c^{\ka-1} \frac{c^2}{z^2}F\left( \frac{1+\ka}{2},1;
\frac{\ka}{2}+1;\frac{c^2}{z^2} \right) \sqrt{\frac{z^2}{c^2}-1} \, ,
\label{Resolvent}
\eeqn
in the one-cut case, for $V(\la)=g |\la|^\ka$.
For any real positive $\ka$, 
the spectral density eq. (\ref{rhofreudF2}) is then recovered by means of 
$2 \pi i \rho(\la)=G(\la-i0)-G(\la+i0)$, $\la \in [-c,c]$ (after applying an 
identity for the Hypergeometric 
function corresponding to the  transformation $\frac{c^2}{z^2} 
\to 1-\frac{z^2}{c^2}$). The normalization condition is 
 readily obtained from the large-$z$ asymptotic behavior of eq. 
(\ref{Resolvent}). In fact by using that $F(\alpha,\beta,\gamma,0)=1$
 one has $G(z) \sim  g H c^{\ka}/z = 1/z$ which is  the previously stated 
formula $g H c^{\ka}=1$.\\

In the remaining part of this section we obtain the spectral density by 
using another technique, 
the method of orthogonal polynomials. It will turn out to be very useful 
also in the next section
 where we prove the universality of all 
microscopic correlation functions away from criticality. 
Let us begin by repeating a few well known formulas \cite{Mehta}
in order to fix our 
notation. We choose a set of orthonormal polynomials,
\beq
\int_{-\infty}^\infty d\la\, w(\la)
 P^{(2N_f,\al)}_n(\la)\,P^{(2N_f,\al)}_m(\la)= \, \delta_{nm}\ ,
\label{OP}
\eeq
where the weight function is taken from eq. (\ref{Z}):
\beq
w(\la) \ \equiv\ |\la|^{2\al}
\prod_{f=1}^{N_f}\left(\la^2\ +\ m_f^2\right)\, e^{-N V(\la) } \ . 
\label{weight}
\eeq
The $k$-point correlation function of the eigenvalues,
\beq
R^{(2N_f,\al)}_N(\la_1,\ldots,\la_k) 
\ \equiv\ \frac{N!}{(N-k)!}
\int_{-\infty}^\infty \prod_{i=k+1}^N\! d\la_i
\  \prod_{j=1}^N w(\la_j)\ 
|\Delta(\la)|^2 \ ,
\label{Rkdef}
\eeq
defined as the $(N-k)$-fold integral over the integrand of eq. 
(\ref{Z}), can be expressed through the 
kernel of the orthogonal polynomials \cite{Mehta}
\beq
R^{(2N_f,\al)}_N(\la_1,\ldots,\la_k) 
\ =\ \frac{N!}{(N-k)!}\det_{1\leq i,j\leq k}
\left[K^{(2N_f,\al)}_N(\la_i,\la_j)\right] \ .
\label{MM}
\eeq 
The kernel is given by
\beqn
K^{(2N_f,\al)}_N(\la,\eta) &=&
w(\la)^{\frac12}w(\eta)^{\frac12}\,
\sum_{i=0}^{N-1} P^{(2N_f,\al)}_i(\la)P^{(2N_f,\al)}_i(\eta)\ \nonumber \\
&=& w(\la)^{\frac12}w(\eta)^{\frac12}\ r_N 
\frac{P^{(2N_f,\al)}_{N}(\la)P^{(2N_f,\al)}_{N-1}(\eta) -
P^{(2N_f,\al)}_{N}(\eta)P^{(2N_f,\al)}_{N-1}(\la)}{\la-\eta} \ ,
\label{kernel}
\eeqn
where in the last step we have used the Christoffel-Darboux identity. It 
follows from a three step recursion relation, which any set of 
orthogonal polynomials obeys. 
For any symmetric potential $V(\la)$, polynomial or not, it reads
\beq
\la P^{(2N_f,\al)}_n(\la)\ = \ r_{n+1}P^{(2N_f,\al)}_{n+1}(\la) 
\ + \ r_n P^{(2N_f,\al)}_{n-1}(\la) \ ,
\label{rec}
\eeq
under the assumption that the integrals over the
$P^{(2N_f,\al)}_n(\la)$ exists.  The recursion coefficients $r_n$ are
determined from the so-called string equation
\beq
n\ =\ -r_n \int_{-\infty}^\infty d\la\, w'(\la)
 P^{(2N_f,\al)}_{n}(\la)\,P^{(2N_f,\al)}_{n-1}(\la) \ .
\label{string}
\eeq
In the case of $\al=N_f=0$, which is all we need in order to determine the 
macroscopic spectral density, one can show that by setting 
$r_N=2c$ eq. (\ref{string}) is 
equivalent to the normalization condition of the spectral density arising 
from the saddle-point analysis (e.g. see eq. (\ref{BC})).

In order to determine the spectral density one has to consider also 
differentiation of orthogonal polynomials.
Here we will use the Shohat method, which is reviewed in great detail 
in \cite{KFrev}. It consists in a particularly useful rewriting of 
$P_n'(\la)$, 
which is in general a linear combination of all lower polynomials, in terms 
of two auxiliary functions  $A_n(\la)$ and $B_n(\la)$.

For the rest of this section we 
will restrict ourselves to the massless case, $N_f=0$, (and suppress the 
index). The massive case with $N_f\neq0$ will be treated later in Section  
\ref{univ}. Furthermore we absorb the massless prefactor $|\la|^{2\al}$ 
from the weight function into the potential
\beq
V_\al(\la) \ =\  V(\la) - \frac{2\al}{N}\ln|\la| \, .
\label{Va}
\eeq
Defining the auxiliary 
functions\footnote{Compared to \cite{KFrev} we have
already used the symmetry of the potential (see also appendix A there).}
\beqn
A_n(\la)&=& N~ r_n\int_{-\infty}^{\infty}\! dt~ 
w(t)\frac{t V_{\al}'(t) - \la V_{\al}'(\la)}{t^2-\la^2}
P^{(\al)}_n(t)^2 ~,
\label{A}\\
B_n(\la)&=&N~ r_n\int_{-\infty}^{\infty}\! dt~ 
w(t)\frac{\la V_{\al}'(t) - t V_{\al}'(\la)}{t^2-\la^2}
P^{(\al)}_n(t)P^{(\al)}_{n-1}(t) \ ,
\label{B}
\eeqn
it is easy to show using the Christoffel-Darboux identity that the 
following relation for differentiation 
holds for any finite $n$ and for any potential $V_\al(\la)$
\beq
\frac{d \ P_n^{(\al)}(\la)}{d  \la} ~\equiv~ A_n(\la)P^{(\al)}_{n-1}(\la) ~-~
                       B_n(\la)P^{(\al)}_{n}(\la) \, .
\label{P'}
\eeq
Furthermore these functions obey the following identity at any finite $n$ 
\cite{KFrev}:
\beq
B_n(\la) + B_{n-1}(\la) ~+~  N~ V_{\al}'(\la) 
~=~ \frac{\la}{r_{n-1}}A_{n-1}(\la)~.
\label{ABV}
\eeq  
The relation (\ref{P'}) 
will be the key to derive an exact differential equation for
the wave functions
\beq
\psi^{(\al)}_{n}(\la)\ \equiv\ w(\la)^{\frac12} P^{(\al)}_{n}(\la)
\label{psi}
\eeq
in the next section and to prove universality.

Now we are using eq. (\ref{P'}) to 
derive the spectral density for the general potential 
defined in eq. (\ref{pot}). From eqs. (\ref{MM}) and (\ref{kernel}) the 
one-point function or spectral density is simply given by a single kernel at 
equal arguments:
\beq
R^{(\al)}_N(\la) \ =\ w(\la)\ r_N \left( 
{P^{(\al)}_{N}}'(\la)P^{(\al)}_{N-1}(\la) -
{P^{(\al)}_{N}}'(\la)P^{(\al)}_{N-1}(\la) 
\right) \ .
\eeq
Using the recursion relation (\ref{rec}) and the identities (\ref{P'}) and 
(\ref{ABV}) we arrive at \cite{KFrev}
\beqn
R^{(\al)}_N(\la) &=& r_N \left[ A_N(\la)\psi^{(\al)}_{N-1}(\la)^2
+\frac{r_N}{r_{N-1}}A_{N-1}(\la)\psi^{(\al)}_{N}(\la)^2 \right.\nonumber \\
&&-\left.\left( \frac{\la}{r_{N-1}} A_{N-1}(\la) 
 +B_{N}(\la) -B_{N-1}(\la)  \right) 
\psi^{(\al)}_{N}(\la)\psi^{(\al)}_{N-1}(\la)
\right] \ ,
\label{RN}
\eeqn
which is exact for any finite $N$. We will now take the macroscopic
large-$N$ limit where we smooth the oscillations of the correlators.
By considering smoothed moments of the orthogonal polynomials at
large-$N$ the following result has been shown in \cite{KFrev}. For any
set of orthogonal polynomials of a given measure $w(\la)$ which first
leads to a large-$N$ spectral density with single interval support
$\si$ and second has recursion coefficients obeying the condition
$r_{N\pm1,\pm2,\ldots}\to r_N=2c$ when $N \to \infty$, it holds
\beqn
\overline{\psi^{(\al)}_N(\la)^2}  &=& \frac{1}{\pi} \frac{1}{\sqrt{c^2-\la^2}} 
\ \theta(c^2-\la^2) \ , \nonumber \\
\overline{\psi^{(\al)}_N(\la)\psi^{(\al)}_{N-1}(\la)} &=& \frac{1}{c\pi} 
\frac{\la}{\sqrt{c^2-\la^2}} \ \theta(c^2-\la^2) \ .
\label{psibar}
\eeqn
Here the bar denotes the smoothed large-$N$ limit.  Let us stress that
the derivation of eq.  (\ref{psibar}) in the Appendix A of
\cite{KFrev} is purely algebraic.  No assumption about the analytic
structure of the spectral density in the complex plane has been made,
in particular no square-root behavior has been assumed.  Using these
results we can take the same smoothed large-$N$ limit for the
functions $A_N(\la)$ and $B_N(\la)$ defined in eqs. (\ref{A}) and
(\ref{B}). Smoothing inside the integrand we obtain
\beqn
A(\la)\ \equiv\ \frac{1}{N}\overline{A_N(\la)}&=& \frac{c}{\pi} 
\int_{0}^{c}\! dt~ \frac{t V'(t) - \la V'(\la)}{t^2-\la^2}
\ \frac{1}{\sqrt{c^2-t^2}} ~,
\label{Abar}\\
B(\la)\ \equiv\ \frac{1}{N}\overline{B_N(\la)}&=& \frac{1}{\pi}  
\int_{0}^{c}\! dt~ \frac{\la V'(t) - t V'(\la)}{t^2-\la^2}
\ \frac{t}{\sqrt{c^2-t^2}} ~.
\label{Bbar}
\eeqn
Note that compared to eqs. (\ref{A}) and (\ref{B}) we have dropped the
index $\al$ in the potential $V(\la)$ or in other words the term $2\al
\ln|\la| /N $ from eq.  (\ref{Va}). This can be seen as
follows. Because of parity it drops out in $A_N(\la)$ already at
finite $N$. In $B_N$ one can split \cite{KFrev}
$B_N(\la)=B_{N,~reg}(\la)+ (1-(-1)^N)\al/\la$ where the first term
only contains $V$ instead of $V_\al$. The latter term obviously has no
smooth limit but it is suppressed by $1/N$ in eq. (\ref{Bbar}).

We can now apply the smooth limit to the density eq. (\ref{RN}) itself
and we obtain
\beq
\rho(\la)\ \equiv\ \frac1N \overline{R^{(\al)}_N(\la)}\ =\ \frac{1}{c\pi}
A(\la) \sqrt{c^2-\la^2} \ ,\ \ \ \ \la\ \in [-c,c]\ .
\label{rhosmooth} 
\eeq
In particular it holds that
\beq
\rho(0) \ =\ \frac{1}{\pi} \ A(0)\ .
\label{rhoA}
\eeq
Eq. (\ref{rhosmooth}) together with eq. (\ref{Abar}) constitute our
main result of this section. Let us stress that it does not
necessarily imply that $\rho(\la)$ has a simple square-root cut as a
function in the complex plane. The function $A(\la)$ having
nonpolynomial parts inside the potential may precisely cancel the
factor $\sqrt{c^2-\la^2}$, as we shall see in a simple example now.
Let us choose again the nonpolynomial potential
\beq
V_{Freud}(\la)=g |\la|^\ka \ , \ \ \ 1< \ka\in \mathbb{R}\ .
\eeq
From eq. (\ref{Abar}) we obtain for the auxiliary function 
\beqn
A(\la) &=& \frac{c}{\pi}\ g\ka\int_{0}^{c}\! dt~
\frac{t^\ka - \la^\ka}{t^2-\la^2}
\frac{1}{\sqrt{c^2-t^2}} \nonumber\\
&=& cg\ka\left( \frac{\la^{\ka-1}}{2\sqrt{c^2-\la^2}}
-\frac{c^\ka H}{\ka\la^2}
F\left(1,\,\frac12+\frac{\ka}{2};\,1+\frac{\ka}{2};\,\frac{c^2}{\la^2}\right)
\right). 
\label{preA}
\eeqn
In order to be able to expand the density around the origin and to see
the square root cancelling in eq. (\ref{rhosmooth}) we still have to
use some identities for Hypergeometric function (and to analytically
continue to $\la^2<c^2$)
\beqn
\frac{1}{\la^2}
F\left(1,\,\frac12+\frac{\ka}{2};\,1+\frac{\ka}{2};\,\frac{c^2}{\la^2}\right)
&=& 
\frac{1}{\sqrt{\la^2-c^2}}\, \frac{1}{\la} 
F\left(\frac{\ka}{2},\,\frac12;\,1+\frac{\ka}{2};\,\frac{c^2}{\la^2}\right)
\\
&=& \frac{1}{\sqrt{\la^2-c^2}} \left( 
\frac{\Gamma(1+\frac{\ka}{2})\Gamma(\frac12-\frac{\ka}{2})}{\sqrt{\pi}}
e^{i\frac{\pi}{2}\ka} \frac{\la^{\ka-1}}{c^\ka} \right.
\nonumber\\
&&+\left.
\frac{\Gamma(1+\frac{\ka}{2})\Gamma(\frac{\ka}{2}-\frac12)}{
\Gamma(\frac{\ka}{2})\Gamma(\frac{\ka}{2}+\frac12)} \frac{i}{c}
F\left(\frac12,\,
\frac12-\frac{\ka}{2};\,\frac32-\frac{\ka}{2};\,\frac{\la^2}{c^2}\right)
\right). \nonumber
\eeqn
Inserting this back into eq. (\ref{preA}) we obtain with eq. (\ref{rhosmooth})
the final expression for our density as given in eq. (\ref{rhofreudF}):
\beq
\rho_{Freud}(\la)\ =\ \frac{g \ka H c^{\ka-1}}{\pi}\left(
\frac{1}{2H}\tan\left[\frac{\ka\pi}{2}\right] 
\left(\frac{\la}{c}\right)^{\ka-1}
+\frac{1}{\ka-1}F\left(\frac12,\,
\frac12-\frac{\ka}{2};\,\frac32-\frac{\ka}{2};\,\frac{\la^2}{c^2}\right)
\right).
\eeq
This solution for the density we have derived can be seen to be
equivalent to the first representation given in
eq. (\ref{rhofreudF2}), by using the identity relating Hypergeometric
functions with argument $z$ and $1-z$ respectively\footnote{When $\ka$
is not an odd integer number. For $\ka$ an odd integer number the
formula is slightly more complicated \cite{abramo}. However, formula
(\ref{rhofreudF}) is completely fine here 
because as we will see in Section \ref{multi},  odd integers $\ka$ do not 
affect the issue of multicriticality we are considering.
}.  The advantage of eq. (\ref{rhofreudF}) is
that we can immediately read off the expansion of the density at the
origin. For noninteger $\ka>1$ it consists of a single term $\sim
\la^{\ka-1}$ plus a power series in $\la^2$.  It is the first term
which will be responsible for the new critical behavior to be
discussed in Section \ref{multi}.


\sect{Microscopic universality}\label{univ} 

In this section we prove that in the microscopic large-$N$ scaling
limit at the origin all correlation functions of our model (\ref{Z})
are universal, as long as the couplings in the potential $V(\la)$
eq. (\ref{pot}) are generic (=noncritical) and we stay in the phase of
a single interval support. They belong to different universality
classes depending on the number of massless and massive flavors, $\al$
and $N_f$ respectively, as well as on the rescaled masses $m_f$. The
universal parameter will be $\rho(0)$ which encodes the influence of
all the coupling constants in the potential via eq. (\ref{rhoA}). Here
we will make use of the powerful results on universality for
polynomial potentials which have been obtained for massless
\cite{ADMN,KF97,ADMNII} 
and massive \cite{DN,DNII,AK} unitary and chiral unitary matrix
models. Our extension to nonpolynomial potentials works as follows.
We start with the unitary model without massive flavors, $N_f=0$,
resuming the method of orthogonal polynomials used in the last
section. A differential equation for the asymptotic of the polynomials
will be derived for arbitrary $\al>-1$ following \cite{KF97}. Once we
have established massless universality we can use the method of
\cite{AK} to extend it to an arbitrary number of massive flavors
$N_f$.  In contrast to the chiral case no restriction on $N_f$ to be
even is needed here, as has been noted already in the extension to the
non Hermitian case \cite{A01}.  In a final step we use a direct
relation between the polynomials of the unitary and the chiral unitary
ensemble \cite{ADMNII} to prove universality also in the chiral case,
for arbitrary $\al$ and $N_f$.

\begin{figure}[h]
\centerline{
\epsfig{figure=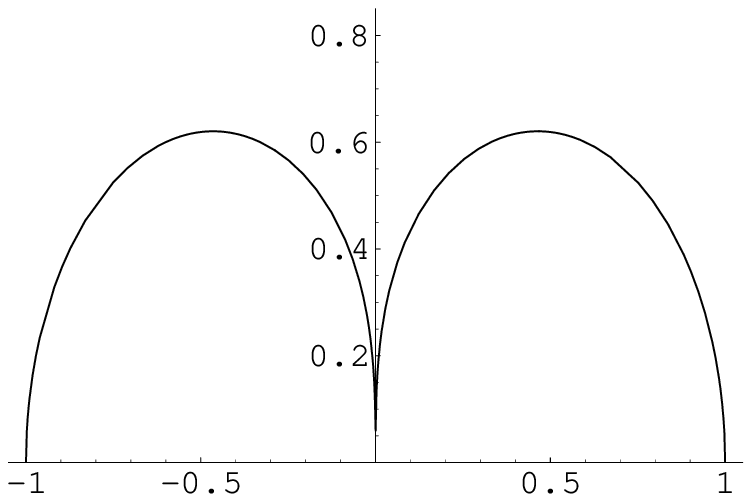,width=20pc}
\epsfig{figure=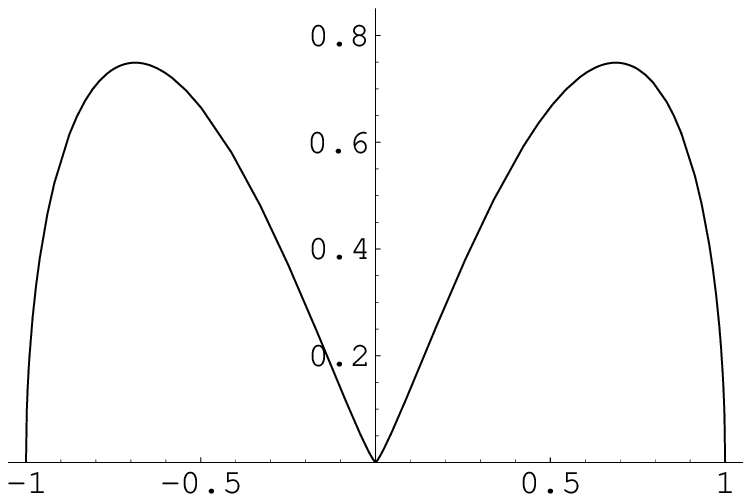,width=20pc}
\put(-110,145){$\rho(\la)$}
\put(-355,145){$\rho(\la)$}
\put(-10.1,0.1){$\la$}
\put(-254,0.1){$\la$}
}
\caption{
The multicritical density for $1<\ka=\frac{4}{3}<2$ (left) and 
for $2<\ka=\sqrt{5}<3$ (right).
}
\label{rhocrit}
\end{figure}

Let us start with the unitary case and $N_f=0$ as in the end of the
last section. Following \cite{KFrev}, eq. (\ref{P'}) for
differentiation on the polynomials can be used together with the
recursion relation (\ref{rec}) and the identity (\ref{ABV}) to derive
an exact second order differential equation for the wave functions
$\psi_n^{(\al)}(\la)$ eq. (\ref{psi}) for any finite $n$ (for the
present form see \cite{ADMNII}):
\beq
\psi^{(\al)}_n(\la)'' - F_n(\la)\psi_n^{(\al)}(\la)' + G_n(\la)
\psi_n(\la) ~=~ 0 ~,
\label{diff}
\eeq
where
\beqn  
F_n(\la) &\equiv& \frac{A_n'(\la)}{A_n(\la)} \ ,
\nonumber \\
G_n(\la) &\equiv& \frac{r_n}{r_{n-1}}A_n(\la)A_{n-1}(\la) 
                -\left(B_n(\la)+\frac{N}{2}V_{\al}'(\la)\right)^2
               \nonumber\\
                &&+\left(B_n(\la)+\frac{N}{2}V_{\al}'(\la)\right)'
                -\frac{A_n'(\la)}{A_n(\la)}
                \left(B_n(\la)+\frac{N}{2}V_{\al}'(\la)\right) ~.
\label{G}
\eeqn
Imposing a smooth limit for the recursion coefficients $r_N$ as well
as for $A_N(\la)$ and the regular part of $B_N(\la)$, the function
$G_N(\la)$ can be entirely expressed in terms of the limiting function
$A(\la)$, due to the identity (\ref{ABV}). We obtain
\cite{ADMNII}
\beq
\psi^{(\al)}_N(\la)'' - \frac{A'(\la)}{A(\la)} \psi_N^{(\al)}(\la)'  \ +
\left( N^2 A^2(\la)\left(1-\frac{\la^2}{c^2}\right)+
\frac{(-1)^N\al-\al^2}{\la^2}+(-1)^N\frac{\al A(\la)'}{\la A(\la)}
\right)\psi_N(\la) ~=~ 0 ~.
\label{diffN}
\eeq
The microscopic large-$N$ limit at the origin is defined\footnote{For
$1<\ka<2$ the slope of $\rho(\la)$ is infinite at the origin (see
Figure
\ref{rhocrit}) and thus the eigenvalues have to be unfolded before
rescaling. We thank K. Splittorff for raising this point.} 
as 
\beq
\xi\ =\ N\la \ ,
\label{soft}
\eeq
where $\xi$ is kept fixed at large-$N$.  This rescaling is the only
consistent way to obtain a finite large-$N$ differential equation from
eq. (\ref{diffN}) and it agrees with the rescaling in the pure Freud
case obtained in \cite{Pastur,CWK,FKY,Ken} when translating to their
conventions.  Eq. (\ref{soft}) holds despite the fact that $A'(\la)$
diverges $\sim\la^{\ka-2}$ for $1<\ka<2$. Even in that case the
logarithmic derivative of $A(\la)$ (as well as $\la^2A(\la)^2$) is
suppressed and we arrive at
\beq
\psi_N(\xi)'' + \left( A(0)^2 
        +\frac{(-1)^N\al-\al^2}{\xi^2}\right) \psi_N(\xi) ~=~ 0 ~.
\label{diffsoft}
\eeq
This equation of Bessel type entirely determines the asymptotic
large-$N$ behavior of the wave function, distinguishing between even
and odd polynomials. Moreover it is universal since it only depends on
the potential $V(\la)$ eq. (\ref{pot}) through the universal parameter
$A(0)=\pi \rho(0)$. We refer to \cite{ADMN,KF97} for the explicit
solution of the differential equation as we are only interested in
proving an extension of universality to nonpolynomial potentials
here. Similarly we refer to \cite{ADMN} for the correlation functions,
where detailed expressions can be found.  We only wish to stress here
that from eq. (\ref{diffsoft}) the rescaled microscopic kernel
\beq
K^{(\al)}_S(\xi,\zeta) \ \equiv \ 
\lim_{N\to\infty} \frac1N K_N^{(\al)}
\left(\frac{\xi}{N},\frac{\zeta}{N}\right)
\label{microkernel}
\eeq
follows as it can be entirely expressed in terms of the asymptotic
wave functions (see eq. (\ref{kernel})). The microscopic correlation
functions defined as
\beq
\rho^{(\al)}_S(\xi_1,\ldots,\xi_k) \ \equiv\ 
\lim_{N\to\infty} \frac{1}{N^k} R^{(\al)}_N
\left(\frac{\xi_1}{N},\dots,\frac{\xi_k}{N}\right)
\label{microRk}
\eeq
can then be obtained from the rescaled version of eq. (\ref{MM}).

Up to now we have determined all microscopic correlation functions
from eq. (\ref{diffsoft}) for an arbitrary $\al>-1$ and proved their
universality.  In the next step we reintroduce the mass terms from
eq. (\ref{Z}) and determine their universal correlations. Following
\cite{AK} these masses can be put into the Vandermonde determinant
leading to a relation between the massive $k$-point correlator and the
massless $(k+N_f)$-point correlator analytically continuated in some of
the arguments.

To be more precise we can rewrite the mass terms times the Vandermonde
in the integrand of eq. (\ref{Z}) as
\beq
\prod_{j=1}^N\prod_{f=1}^{N_f} (\la_j-im_f)(\la_j+im_f)
\ |\Delta_N(\la_1,\ldots,\la_N)|^2
\ =\ \frac{|\Delta_{N+N_f}(\la_1,\ldots,\la_N,im_1,\ldots,im_{N_f})|^2}
{|\Delta_{N_f}(im_1,\ldots,im_{N_f})|^2} \ .
\label{delta}
\eeq
Here we have explicitly given all the arguments of the Vandermonde
determinant.  Because of the pairing of the $2N_f$ masses in pairs
$\pm i m_f$ coming from QCD3 we do not need to impose an extra
degeneracy to the power of the Dyson index $\beta=2$ as it was done in
the chiral case \cite{AK}. This observation valid for the unitary case
was made in \cite{A01}.  Inserting eq. (\ref{delta}) into the
definition (\ref{Rkdef}) we see that the right hand side is
proportional to a massless $k+N_f$-point function, with no mass
terms. Inserting all normalization factors we arrive at
\cite{A01}\footnote{To be precise the relations holds for finite-$N$
only in the normalization of \cite{AK}. At large-$N$ this becomes
irrelevant.}
\beq
R_N^{(2N_f,\al)}(\la_1,\ldots,\la_k) \ =\
\frac{R_{N+N_f}^{(0,\al)}(\la_1,\ldots,\la_k,im_1,\ldots,im_{N_f})}
{R_{N+N_f}^{(0,\al)}(im_1,\ldots,im_{N_f})} \ .
\label{master}
\eeq
Since for the right hand side we have already shown its universality
in the microscopic limit the same statement holds for the left hand
side.  Here the masses have to be rescaled the same way as the
eigenvalues, keeping $\mu_f\equiv N m_f$ fixed. We have thus
calculated the massive correlation functions and at the same time
given a proof for their universality. Explicit expressions in
different equivalent forms can be found in \cite{DNII,Sz}. We have
thus not only extended the universality proof of \cite{DNII} to
potentials of the form in eq.  (\ref{pot}) but also considerably
shortcut proof and calculations.

After having completed the universality proof for the full model
eq. (\ref{Z}) it remains to repeat the analysis for the corresponding
chiral model.  The partition function can be written in two ways:
\beqn
{\cal Z}_{chiral}^{(N_f,\al)}(\{m_f\}) 
&=& \int_0^\infty \prod_{i=1}^N 
\left( d\la_i\  \la_i^{\al}
\prod_{f=1}^{N_f}\left(\la_i\ +\ m_f^2\right)\, 
e^{-N V_{chiral}(\la_i) }\right)
|\Delta(\la)|^2 \   \nonumber \\
&=& \int_{-\infty}^\infty \prod_{i=1}^N 
\left( d\la_i\  |\la_i|^{2\al+1}
\prod_{f=1}^{N_f}\left(\la_i^2\ +\ m_f^2\right)\, 
e^{-N V_{chiral}(\la_i^2) }\right)
|\Delta(\la^2)|^2 \ ,  
\label{Zchiral}\\
V_{chiral}(\la^2) &\equiv& 
\sum_{j=1}^{d_p} \frac{g_{2j}}{j} \la^{2j} \ +\ 
         \sum_{j=1}^{d_{np}} 2{h_j} |\la|^{\ka_j} \ \ \ ,
\label{potch}
\eeqn
where in the second line we have simply changed variables $\la \to
\la^2$.  In order to calculate the correlation functions we introduce
again orthogonal polynomials in analogy to eq. (\ref{OP}):
\beqn
\delta_{nm}&=&
\int_0^\infty d\la\, \la^\al
\prod_{f=1}^{N_f}(\la+m_f^2)
\ e^{-N V_{chiral}(\la) }
 P_{n,chiral}^{(N_f,\al)}(\la)\,P_{m,chiral}^{(N_f,\al)}(\la) 
\label{OPch}\\
&=&\int_{-\infty}^\infty dz\, |\la|^{2\al+1}
\prod_{f=1}^{N_f}(\la^2+m_f^2)
\ \mbox{e}^{-N V_{chiral}(\la^2) }
 P_{n,chiral}^{(N_f,\al)}(\la^2)\,P_{m,chiral}^{(N_f,\al)}(\la^2)\ .\nonumber
\eeqn
By comparing to the unitary orthogonal polynomials eq. (\ref{OP}) we
can immediately read off that the two sets are related. This relation
has been already found in \cite{ADMNII} and trivially extends to the
massive case
\cite{ADIII}:
\beq
P_{2N}^{(2N_f,\al+1/2)}(\la)\ =\ P_{N,chiral}^{(N_f,\al)}(\la^2)\ ,
\label{OPrel}
\eeq
where we also have to identify the corresponding potentials
\beq
2 V(\la) \ \equiv\   V_{chiral}(\la^2)\ .
\label{pots}
\eeq
In other words the even subset of the polynomials of the Unitary
Ensemble at shifted $\al+\frac12$ is sufficient to construct the full
set polynomials of the chiral ensemble. The same is true for the wave
functions.  Since we have started with a real $\al$ this shift is
possible. We can thus borrow the full machinery for the asymptotic of
the polynomials in the unitary case, including universality.  We have
shown already that through eq. (\ref{diffsoft}) the asymptotic of the
wave functions in the massless unitary case are the same for
polynomial and nonpolynomial potentials and that they are
universal. Through eq. (\ref{OPrel}) the same statements immediately
translate to the wave functions of the massless chiral ensemble and
thus through eqs. (\ref{kernel}) and (\ref{MM}) to all massless
correlation functions. For the massive correlations we have to invoke
another relation since for the Unitary Ensemble we did not determine
the massive orthogonal polynomials. We directly determined the massive
correlation functions through eq. (\ref{master}) and thus we cannot
use eq. (\ref{OPrel}) to read off the massive chiral orthogonal
polynomials.  Fortunately the corresponding massive kernels of the two
ensembles, chiral and non chiral, can be directly related as well by
using eq. (\ref{OPrel}) \cite{ADIII}:
\beq
K_{N, chiral}^{(N_f,\al)}(\la^2,\eta^2) \ =\ 
\frac12 \left( 
K_{2N}^{(2N_f,\al+1/2)}(\la,\eta) \ +\ K_{2N}^{(2N_f,\al+1/2)}(-\la,\eta)  
\right)\ .
\label{kernelrel}
\eeq
The massive kernel of the Unitary Ensemble on the right hand side
follows from the massive two-point function that we have already
determined in eq. (\ref{master}). Inspecting eq. (\ref{MM}) we see
that it follows from
\beq
K_{N}^{(2N_f,\al)}(\la,\eta) \ =\ \frac1N
\sqrt{-R_{N, conn}^{(2N_f,\al)}(\la,\eta)}
\ ,
\eeq
where the connected part of the correlation function is defined as 
\beq
R_{N, conn}^{(2N_f,\al)}(\la,\eta)\ \equiv\ 
R_{N}^{(2N_f,\al)}(\la,\eta)-R_{N}^{(2N_f,\al)}(\la)R_{N}^{(2N_f,\al)}(\eta)
\ .
\eeq 
Using again eq. (\ref{MM}) we have finally also determined all massive
correlation functions of the chiral ensemble. Our consideration
establishes an alternative proof to the original one in \cite{DN}, in
a considerably simpler way. This is the gain we made by first solving
the nonchiral ensemble.  Let us stress that in contrast to \cite{AK}
we have no restriction on the flavor number $N_f$.

Finally we mention that the identification of the potentials,
eq. (\ref{pots}) is particularly useful when studying the
multicritical point as it will be done in the next section. We can
therefore restrict ourselves again to the unitary case.


\sect{New multicritical behavior at the origin}\label{multi}

Multicritical points occur when the spectral density $\rho(\la)$ develops 
additional zeros inside (or at the edge of) the support of the eigenvalues. 
Since in this section 
we are interested in applications to the chiral phase transition in 
QCD, with the chiral condensate being proportional to $\rho(0)$, we restrict 
ourselves to additional zeros at the origin of the spectrum. 
The multicritical points can also be thought of the support merging together 
from two 
(or more) segments. In fact for the question of the order of the phase 
transition the free energy and its derivatives have to be calculated 
from both sides and compared. In general these transitions are of third order 
which also holds for additional zeros developing elsewhere on the support. 

Before defining the new class of critical potentials let us briefly
review what is known about multicritical models with polynomial
potentials.  Since we only study the behavior of the macroscopic
density for the degree of criticality we can set again $N_f=\al=0$
since it does not depend on these parameters.  From eq. (\ref{Abar})
we deduce that for polynomial potentials the density (\ref{rhosmooth})
consists of an even polynomial times a square root $\sqrt{c^2-\la^2}$.
Expanding $\rho(\la)$ at the origin and tuning the couplings for
$\rho(0)$ to vanish we can only achieve that $\rho(\la=0)\sim
\la^{2m}$ for $m\in\mathbb{N}$. For the $m$th multicriticality we have
to tune precisely $m$ coupling constants. In \cite{MC} a set of
minimal potentials was given.  The corresponding $m$th multicritical
correlation functions were determined in \cite{ADMNII} and shown to be
universal.  We will find a similar situation here (although we will
not touch the issue of universality), namely for a spectral density
vanishing like a real power $\rho(\la=0)\sim \la^{\ka-1}$, with
$2m-2<\ka-1<2m$ we have to tune $m$ coupling constants.

To see this let us start with the simplest example, a Gaussian plus a single 
real positive power:
\beq
V(\la) = \frac12 g_2 \la^2 \ +\ g |\la|^{\ka} \ . 
\label{VGF}
\eeq
Instead of the solution eq. (\ref{rhoGF}) we use the equivalent representation 
for the integral, eq. (\ref{rhofreudF}):
\beqn
\rho(\la) &=& 
\frac{g \ka}{2\pi}\tan\left[\frac{\ka\pi}{2}\right] \la^{\ka-1}\ 
+ \ 
\frac{g \ka H c^{\ka-1}}{\pi(\ka-1)}
F\left(\frac12,\,
\frac12-\frac{\ka}{2};\,\frac32-\frac{\ka}{2};\,\frac{\la^2}{c^2}\right)\ 
+\ 
\frac{1}{2\pi} g_2 \sqrt{c^2-\la^2}  ,
\label{rhoGFF}
\eeqn
where we now have to restrict to $1<\ka\neq 3,5,7,\ldots$ . The normalization 
condition eq. (\ref{BC}) which we repeat here for completeness 
fixes one of the coupling constants
\beq
1 \ = \  \frac14 g_2 c^2 + Hg c^\ka \ .
\label{BCr}
\eeq 
When expanding the density at the origin the Hypergeometric function as well 
as the square root both have an expansion in $\la^2$ and we obtain 
\beqn
\rho(\la) &=& 
\frac{g \ka}{2\pi}\tan\left[\frac{\ka\pi}{2}\right] \la^{\ka-1}\ 
+ \ 
\frac{1}{2\pi}\left( 
\frac{2\ka}{\ka-1} g H c^{\ka-1} \ +\ g_2 c
\right)
\ +\ {\cal O}(\la^2)\ .
\label{rhoGFexp}
\eeqn
By tuning the remaining coupling constant we can achieve $\rho(0)=0$ through
\beq
0\ =\ \frac{2\ka}{\ka-1} g H c^{\ka} \ +\ g_2 c^2\ .
\label{multi1}
\eeq
However, only for $\ka<3$ the first term in eq. (\ref{rhoGFexp}) 
with a real power is dominant and we have 
\beq
\rho(\la=0) \ \sim\ 
\frac{g \ka}{2\pi}\tan\left[\frac{\ka\pi}{2}\right] 
\  \la^{\ka-1}
\ +\ {\cal O}(\la^2)
\ \ \ \mbox{for} \ \ 1<\ka<3 \ \ .
\label{ka<3}
\eeq
Consequently the potential eq. (\ref{VGF}) together with the condition
eq. (\ref{multi1}) and the normalization eq. (\ref{BCr}) defines our
first new class (\ref{ka<3}) of multicritical models.  A few comments
are in order. First of all it is easy to see that equations
(\ref{multi1}) and (\ref{BCr}) have a unique solution with $g_2>0,\
g<0$ for $1<\ka<2$ and with $g_2<0,\ g>0$ for $2<\ka<3$. In both cases
the slope of $\rho(\la)$ in eq. (\ref{ka<3}) is positive (infinite or
finite), as it should be.  An example for each of such cases is given
in Figure \ref{rhocrit}.  

For real $\ka>3$ the quadratic term in 
(\ref{rhoGFexp}) is leading and we are back in the class  
$\rho(\la=0)\sim \la^{2m}$ for $m=1$ \cite{ADMNII}.
If we add a higher order term to the polynomial part of the potential, like a 
quartic power for example, we can still achieve that $\rho(0)=0$. There is one 
free parameter left and we remain in the multicritical class of \cite{ADMNII} 
unless we also cancel the $ {\cal O}(\la^2)$ term. In the latter case, which 
we will give as a second example below we may arrive at  higher criticality, 
$\rho(\la=0)\sim \la^{\ka-1}$, with $3<\ka<5$. 
Before doing that let us remark that instead of 
eq. (\ref{VGF}) we could have started with a potential  
$V(\la) = h_1 |\la|^{\ka_1}+h_2 |\la|^{\ka_2}$ instead of perturbing 
around the Gaussian. Going through the derivation again, where we now add two 
densities of the type eq. (\ref{rhofreudF}), we could again achieve that 
$\rho(0)=0$. The leading order term of $\rho(\la)$ would then be to the power 
of $\min\{\ka_1-1,\ka_2-1,2\}$. Hence in 
case of $1<\ka_i<3$ for at least one of 
the $\ka_i$ we would be again in class (\ref{ka<3}), otherwise in class 
$m=1$ of \cite{ADMNII}. Adding higher order terms, polynomial or not, would 
then allow to obtain higher criticality as well. 

In order to keep the algebra simple we make the minimal choice of one real 
power plus a polynomial part for the minimal critical 
potentials. As an example  
for the second multicritical class we thus add a quartic term to eq. 
(\ref{VGF}) and start with 
\beq
V(\la) = \frac12 g_2 \la^2 \ +\ \frac14 g_4 \la^4 \ +\ g |\la|^{\ka} \ . 
\label{VQF}
\eeq
The density can be obtained using eqs. (\ref{rhosmooth}) and (\ref{Abar}) or
by simply adding the well known density of the purely quartic potential 
\cite{BPIZ}
to eq. (\ref{rhoGF}) and normalizing properly.
We obtain 
\beqn
\rho(\la) &=& 
\frac{g \ka}{2\pi}\tan\left[\frac{\ka\pi}{2}\right] \la^{\ka-1}\ 
+ \ 
\frac{g \ka H c^{\ka-1}}{\pi(\ka-1)}
F\left(\frac12,\,
\frac12-\frac{\ka}{2};\,\frac32-\frac{\ka}{2};\,\frac{\la^2}{c^2}\right)
\nonumber\\
&&+\ 
\frac{1}{2\pi}\left( g_2+ \frac12 g_4c^2+g_4\la^2\right)\sqrt{c^2-\la^2} \ , 
\label{rhoQFF}
\eeqn
with the normalization condition
\beq
1 \ = \  \frac14 g_2 c^2\  +\  \frac{3}{16} g_4 c^4\  +\   Hg c^\ka \ .
\eeq
Expanding at $\la=0$ and setting the terms of  ${\cal O}(1)$ and 
${\cal O}(\la^2)$ to zero we obtain
\beq
\rho(\la) \ \sim \  
\frac{g \ka}{2\pi}\tan\left[\frac{\ka\pi}{2}\right] 
\la^{\ka-1}\ +\ {\cal O}(\la^4)
\ \ \ \mbox{for} \ \ 3<\ka<5 \ \ .
\label{ka<5}
\eeq
with conditions
\beqn
0&=& \frac{2\ka}{\ka-1} g H c^{\ka} \ +\ g_2 c^2\ +\ \frac12 g_4c^4\ ,
\nonumber\\
0&=& \frac{\ka}{\ka-3} g H c^{\ka} \ -\ \frac12g_2c^2\ +\ \frac34 g_4c^4\ .
\label{multi2}
\eeqn

After giving two explicit examples we now wish to come to the general
$m$th multicritical density and potential. In order to have the first term in
eq. (\ref{rhoQFF}) dominating for $2m-2<\ka<2m$ we have to add a polynomial
potential with at least $m$ coupling constants in order to cancel the first
$m$ terms in the expansion of the Hypergeometric function there. Generalizing 
the cases eq. (\ref{VGF}) and eq. (\ref{VQF}), hereafter we consider the 
potential 
\beq
V(\la) \ = V_{pol}(\la) +g |\la|^\ka= \ \sum_{j=1}^{m} \frac{g_{2j}}{2j}
\la^{2j} \ +\ g |\la|^\ka \ .
\label{Vmthcrit}
\eeq
The general one-cut solution for the polynomial part is know to be \cite{BZ} 
\beq
\rho_{pol}(\la) =
\frac{1}{2\pi} \sum_{j=1}^m g_{2j} \sum_{k=0}^{j-1} \left(
\begin{array}{c}
2k\\
k
\end{array}
\right) \frac{c^{2k}}{2^{2k}} \la^{2j-2k-2}
\sqrt{c^2-\la^2}  \, ,
\label{Vrhopol}
\eeq
which we rewrite as 
\beq
\rho_{pol}(\la) \ =\ 
\sum_{j=0}^{m-1}a_j \frac{\la^{2j}}{c^{2j}}
\sqrt{1-\frac{\la^2}{c^2}}\ ,
\label{rhopol}
\eeq
with
\beq
\label{defal}
a_j \ \equiv \ \frac{c\, 2^{2 j-1}}{ \pi} \sum_{k=j}^{m-1} g_{2k+2}\left(
\begin{array}{c}
2(k-j)\\
k-j
\end{array}
\right) \left(\frac{c}{2}\right)^{2k} \, , \quad j=0,\ldots,m-1 \ .
\label{gsystem}
\eeq
The full density is obtained by adding  the Freud-type density eq. 
(\ref{rhofreudF}) to eq. (\ref{rhopol}), 
and the normalization condition reads
\beq
1\ =\ gHc^\ka + c \pi
\sum_{j=0}^{m-1}
a_j\ 
\frac{(2j-1)!!}{2^{j+1}(j+1)!} \, ,
\label{multinorm}
\eeq
with $(-1)!! \equiv 1$. 
When expanding the Hypergeometric function at 
$\la=0$ we further obtain $m$ equations by setting the terms 
${\cal O}(1)$, \ldots,  ${\cal O}(\la^{2m-2})$ to zero. 
We find\footnote{The meaningless product $\prod_{l=1}^0$ which appears 
when $j=0$, is set equal to $1$.}
\beq
a_j \ =\ \frac{g\ka H c^{\ka-1}}{\pi (1-\ka)}
\prod_{l=1}^j \frac{(2l-\ka)}{(2l+1-\ka)} 
\ \ , \ j=0,\ldots,m-1 \ .
\label{als}
\eeq
The $m$th multicritical density is thus reading 
\beqn
\rho(\la) &=& 
\frac{g \ka H c^{\ka-1}}{\pi(\ka-1)}\left(
F\left(\frac12,\,
\frac12-\frac{\ka}{2};\,\frac32-\frac{\ka}{2};\,\frac{\la^2}{c^2}\right)
\ -\ \sum_{j=0}^{m-1}\prod_{l=1}^j \frac{(2l-\ka)}{(2l+1-\ka)} \ 
\frac{\la^{2j}}{c^{2j}}\sqrt{1-\frac{\la^2}{c^2}}
\right)
\nonumber\\
&&+\ 
\frac{g \ka}{2\pi}\tan\left[\frac{\ka\pi}{2}\right] \la^{\ka-1}\ \ , 
\ \ \ \ \ \ \ \ \ \ \ \ 2m-1<\ka<2m+1\ ,
\label{rhomulti}
\eeqn
and it behaves like $\rho(\la=0)\sim \la^{\ka-1}+{\cal O}(\la^{2m})$.
The critical values of the $m+1$ coupling constants
$g^c_{2j}$ and $g^c$ in the potential eq. (\ref{Vmthcrit}) 
can be obtained by iteratively solving the
linear set of $m+1$ equations given by eq. (\ref{gsystem}) and
eq. (\ref{multinorm}), which is already of triangular form.
Actually there exists a much more compact form for
eq. (\ref{rhomulti}). 
It can be obtained by iteratively using the identity 
\beq
\frac{(\ka-2)}{(\ka-3)}\ z^2\ 
F\left(\frac12,\,
\frac32-\frac{\ka}{2};\,\frac52-\frac{\ka}{2};\,z^2\right) 
\ =\  F\left(\frac12,\,
\frac12-\frac{\ka}{2};\,\frac32-\frac{\ka}{2};\,z^2\right) 
\ -\ \sqrt{1-z^2} \ ,
\label{identity}
\eeq
which can be derived using standard formulas \cite{abramo} . 
Going back to the first criticality $m=1$, in eq. (\ref{rhoGFF}) the square 
root and the Hypergeometric function can be put together using the identity 
(\ref{identity}) and eq. (\ref{multi1}) (see also eq. (\ref{rhoIId=0})). At 
$m=2$ for eq. (\ref{rhoQFF}) we can do the same. The resulting Hypergeometric 
function can be simplified again together with the term 
$\la^2\sqrt{c^2-\la^2}$, using the identity (\ref{identity}) at $\ka\to\ka-2$.
Proceeding by induction with the general form eq. (\ref{rhomulti}) we arrive 
at our final result:
\beq
\label{finalrho}
\rho(\la) = 
\frac{g \ka}{2\pi}\tan\left[\frac{\ka\pi}{2}\right] \la^{\ka-1}+
\frac{g \ka H c^{\ka-1}}{\pi(\ka-1)}  
\frac{\la^{2m}}{c^{2m}}
F\left(\frac12,\frac12+m -\frac{\ka}{2};
\frac32+m-\frac{\ka}{2};\frac{\la^2}{c^2} \right) 
\prod_{l=1}^m \frac{(2l-\ka)}{(2l+1-\ka)} \, .
\eeq
Here the vanishing 
at $\la=0$ is manifest. Notice that in our parameterization, the end-cut
point $c$ is a given positive real number: that means that in general
one has at most a critical line in the space of the parameters of the
model. This finishes our complete classification of new multicritical
models.\\

We turn now to the microscopic scaling limit of the multicritical models.
While the differential equation (\ref{diff}) together with (\ref{G}) 
for the wave functions $\psi^{(\al)}_n(\la)$ still holds exactly 
for finite $N$ the
large-$N$ limit will change. We first determine the appropriate microscopic
scaling limit that corresponds to eq. (\ref{soft}) at criticality. For that
purpose we look at the large-$N$ version of the differential equation
(\ref{diffN}). Although this equation will acquire additional corrections
at multicriticality, as it happens at multicriticality for polynomial
potentials \cite{ADMNII}, it will be sufficient to derive the correct scaling
limit. Assuming also the relation (\ref{rhosmooth}) to hold, which will again 
have
to be corrected as in \cite{ADMNII}, implies $A(\la)\sim\la^{\ka-1}$ at
$\la=0$. This leads to the following approximate equation at $\la \approx 0$
\beqn
&&\psi^{(\al)}_N(\la)'' - \frac{\ka-1}{\la}\psi_N^{(\al)}(\la)'\ + 
\label{diffnaiv}\\
&&\left(  N^2
\frac{g^2\ka^2}{4} \tan^2\left[\frac{\ka\pi}{2}\right]
\la^{2\ka-2}\left(1-\frac{\la^2}{c^2}\right)+
\frac{(-1)^N\al-\al^2}{\la^2}+(-1)^N\frac{\al (\ka-1)}{\la^2}
\right)\psi_N(\la) ~=~ 0 ~.\nonumber
\eeqn
It is easy to see that in order to obtain a nontrivial large-$N$ 
limit we have to define 
\beq
\xi \ =\ N^{1/\ka}\la
\label{microcrit}
\eeq
as the appropriate microscopic scaling limit at criticality\footnote{ 
It generalizes the scaling behavior found in \cite{BH98} from a different
matrix model: $\xi=N^{(2q+1)/(2q+2)}\la$, $q\in\mathbb{N}$. 
There, multicriticality is reached by adding and tuning an
external matrix to a Gaussian model.}.
The naive
differential equation obtained with the rescaling eq. (\ref{microcrit}) 
then reads 
\beq
\psi^{(\al)}_N(\xi)'' - \frac{\ka-1}{\xi} \psi_N^{(\al)}(\xi)'  \ +
\left( \frac{g^2\ka^2}{4}\tan^2\left[\frac{\ka\pi}{2}\right]\xi^{2\ka-2}
+ \frac{(-1)^N\al\ka-\al^2}{\xi^2}
\right)\psi_N(\xi) ~=~ 0 ~.
\label{diffmicronaiv}
\eeq
This equation is clearly no longer of Bessel type and thus the corresponding 
correlation functions following from it belong to a new class. However, 
it cannot be the correct differential equation because it is not invariant 
under
the following symmetry on wave functions, 
\beq
\psi_{2n}^{(\al+1)}(\la) \ =\ \psi_{2n+1}^{(\al)}(\la)\ .
\label{psirel}
\eeq
This relation has been shown to hold in \cite{ADMNII}, independent of the fact
whether the potential is polynomial or not. 
However, eq. (\ref{diffmicronaiv}) may hold in an approximate sense as in
\cite{ADMNII} when in analogy to there the condition 
$1\ll \frac{g^2\ka^2}{4}\tan^2\left[\frac{\ka\pi}{2}\right]\xi^{2\ka-2}$ 
is satisfied.

In order to derive the correct differential equation at multicriticality we
would have to proceed as in \cite{ADMNII} and analyze the corrections to
eq. (\ref{diffN}) from (\ref{diff}) which are no longer subdominant at
criticality. However, because of the nonpolynomial part of the potential this
is forbiddingly difficult. Following \cite{ADMNII} we first would have to
derive the auxiliary functions $A_n(\la)$ and $B_n(\la)$, eqs. (\ref{A}) and
(\ref{B}) respectively, at finite $n$. Choosing an explicit example for the
potential such as eq. (\ref{VGF}) in the beginning of this section one would
insert it into the definition (\ref{A}) and try to evaluate it at finite
$n$. The obvious way to do so would be to use the recursion relation
eq. (\ref{rec}) and the orthogonality of the polynomials. We arrive
at 
\beq
A_n(\la)\ =\ N r_n\left( g_2 +  \ka\int_{-\infty}^{\infty}\! dt~ 
w(t)\frac{t^\ka - \la^\ka}{t^2-\la^2}
P^{(\al)}_n(t)^2 \right)~.
\label{An}
\eeq
The integral can only be evaluated at large-$N$ using the result
eq. (\ref{psibar}) and not at finite $n$ as we need. 
Let us point out a second difficulty  
in the finite-$N$ analysis of the string equation (\ref{string})
which determines the recursion coefficients $r_n$ as a functions of the
coupling constants. Taking the large-$n$ limit and imposing 
$r_{n\pm1}\to r_n$ we easily obtain   
\beq
\frac{n}{N}\ =\ g_2r_n^2  \ +\ g H (2r_n)^\ka \ ,
\label{stringN}
\eeq
using eq. (\ref{psibar}). At $n=N$ we obtain the normalization
condition eq. (\ref{BCr}) using $2r_N=c$, as we have mentioned already in the
beginning. However, it is unclear how to derive the finite-$n$ version of
eq. (\ref{stringN}) in terms of $r_{n\pm1},\ r_{n\pm2}$, etc. It may happen
that the string equation contains all recursion coefficients from $r_N$ down
to $r_0$.

Let us summarize our findings for the multicritical points at the origin 
obtained so far. We
have shown that new classes of multicritical models exist, with $\rho(\la)\sim
\la^{\ka-1}$ for all real  
$\ka$ with $2m-1<\ka<2m+1$ and
$m\in\mathbb{N}$, giving explicit examples. We have shown that at criticality
the eigenvalues scale with $N^{1/\ka}$ (eq. (\ref{microcrit})) and we
have given an approximate differential equation for the wave functions,
eq. (\ref{diffmicronaiv}). 
Because of the relations (\ref{OPrel}) and (\ref{pots}) to the chiral ensemble
all the results obtained in this section immediately translate to the chiral
matrix model eq. (\ref{Zchiral}). 
In the next section we will investigate if the
nonpolynomial potentials also lead to new critical exponents in the large-$N$
scaling limit at the edge of the spectrum.


\sect{Recovering usual multicriticality at the spectrum edge}\label{softmulti}

In this section we investigate the multicritical points at the endpoint $c$ 
of the 
spectrum. This is particularly interesting since it is known from matrix models
with polynomial potentials that here the free energy displays critical
behavior, ${\cal F}\sim (g-g_c)^{2-\ga_{str}}$, with the critical exponents 
$\ga_{str}=-\frac1m$ coinciding with the minimal models \cite{Kazakov}. 
One might
expect that by adding a nonpolynomial part to the potential these exponents
could be altered, as has been already suggested in \cite{GM}. 
In particular one could hope that because of the presence
of a real parameter, $\ka>1$, these exponents could even be nonrational, thus
representing a one-matrix model representation of such conformal field
theories, supposedly again a nonunitary one. 
However, we will find that this is not true in our case. 
We will first expand the
density around the endpoint $\la=c$ and tune to criticality. It turns out that
additional zeros occur precisely in the same way as in the polynomial case,
with $\rho(\la)\sim (c^2-\la^2)^{\frac32}$ in the simplest case. 
We also calculate the free energy
and its derivatives and find the same critical exponent as in \cite{Kazakov}. 

In order to be most explicit we have chosen to work again with the simplest 
example, a Gaussian plus a real power, eq. (\ref{VGF}).  
Since we find the same phenomenon 
as for the corresponding critical polynomial potential
we do not expect changes when adding higher order terms and tuning to higher
criticality. 

We start by expanding the density eq. (\ref{rhoGFF}) for the potential 
$V(\la) = \frac12 g_2 \la^2+g |\la|^{\ka}$,  eq. (\ref{VGF}), around the
endpoint of support $c$. We find that it has an expansion in powers
$(c^2-\la^2)^{\frac12}$ which immediately follows from eq. (\ref{rhofreudF2})
for its Freud part:
\beq
\rho(\la) \ =\  
\frac{1}{2\pi}\left( g_2 + 2\ka g c^{\ka-2}H\right) (c^2-\la^2)^{\frac12}
\ +\ {\cal O}\left((c^2-\la^2)^{\frac32}\right) \ \ \mbox{at} \ \ 
\la^2\approx c^2 \ .
\label{rhocritsoft}
\eeq
By requiring 
\beq
0\ =\ g_2\ +\ 2\ka g c^{\ka-2}H \ ,
\label{softcrit}
\eeq
we obtain an additional zero at the edge of the spectrum.
Together with the normalization condition for the density, eq. (\ref{BC}) 
$1=\frac14 g_2 c^2+Hg c^{\ka}$, 
this fixes both coupling constants $g_2$ and $g$. 
We thus have for the critical endpoint
$c_c$ in terms of $g_2$:
\beq
c_c^2 \ \equiv\ \frac{4\ka}{g_2(\ka-2)} \ .
\label{c_crit}
\eeq
The fact that the density vanishes with the same power as the density of the
critical quartic potential, does not suffice to conclude that the critical
exponents of the corresponding free energies are the same. The free energy is
given as an integral over the full support and thus may very well behave
differently. We use here the saddle-point approach to compute the free energy
recalling some known facts. First, we calculate the free energy in general
and then switch to the critical point (of its derivatives) which turns out to
coincide with eq. (\ref{softcrit}).  

The planar free energy is given by inserting the solution $\rho(\la)$ of the
saddle-point equation (\ref{SP}) into eq. (\ref{F}). Using that $\rho(\la)$
satisfies (\ref{SP}) (here we put the normalization to unity, $a=1$) and 
eliminating the Lagrange multiplier $K$ we obtain 
\beq
{\cal F}\ =\ \frac12 \int_\si d\la \rho(\la) V(\la) 
-\int_\si d\la\, \rho(\la)\ln|\la| \ .
\label{FSP}
\eeq
Choosing our specific example eq. (\ref{VGF}) 
we insert the density in the form of
eq. (\ref{rhoGF}) into eq. (\ref{FSP}). The following rewriting of
integrals, 
\beq
\int_{-c}^c d\la \int_{|\frac{\la}{c}|}^1 ds 
\ =\ \int_0^1 ds \int_{-s}^s \frac{d\la}{c}  \ ,
\eeq
turns out to be particularly useful and the remaining integrals are
straightforward. We obtain  
\beq
{\cal F}\ =\ \frac{3}{2\ka}\ -\ \frac12\ln\left[\frac{c^2}{4} \right]
\ +\ \frac{(\ka-2)(\ka+1)}{4\ka(\ka+2)}\,g_2 c^2 \ - \
\frac{(\ka-2)^2}{64\ka(\ka+2)}\,g_2^2 c^4 \ .
\label{Fsol}
\eeq
In order to find the critical point and exponent of the free energy we have to
study its derivatives with respect to some coupling constant. Since we do not
yet impose criticality only one of the coupling constants $g_2$ and $g$ is
fixed by the normalization condition eq. (\ref{BC}). We choose to keep the
Gaussian coupling $g_2$ as fixed parameter 
and vary $g$ and thus $c$ as a function of $g$. Its derivative can be
obtained from the normalization condition, eq. (\ref{BC}), keeping $g_2$
fixed:
\beqn
\frac{\partial c^2}{\partial g} 
&=& \frac{-4Hc^\ka}{g_2+2\ka Hg c^{\ka-2}} \ , \nonumber \\
&=& \frac{8Hc^{\ka+2}}{g_2(\ka-2)} \frac{1}{(c^2-c_c^2)} \ ,
\label{dc}
\eeqn
where we have used again  eq. (\ref{BC}) as well as the definition 
eq. (\ref{c_crit}). The derivative has a pole precisely at the critical value 
$c_c$. 
The first derivative of the free energy thus reads
\beqn
\frac{\partial {\cal F}}{\partial g} &=& \frac{-4Hc^\ka}{g_2+2\ka Hg c^{\ka-2}}
\left( -\frac{1}{2c^2} \ +\ \frac{(\ka-2)(\ka+1)}{4\ka(\ka+2)}\,g_2 \ + \
\frac{(\ka-2)^2}{32\ka(\ka+2)}\,g_2^2 c^2 \right), \nonumber \\
&=& \frac{Hc^\ka}{4\ka(\ka+2)}\left(4\ka +(2-\ka)g_2c^2+8\right) \ ,
\label{F'}
\eeqn
where the denominator from  eq. (\ref{dc}) has been canceled due to
eq. (\ref{BC}). Obviously $\dga {\cal F}$ is analytic everywhere 
for $\ka>1$ 
and at the critical point for the density eq. (\ref{softcrit}) we can
just set $c=c_c$. The second derivative can be calculated in a similar way and
we obtain 
\beq
\frac{\partial^2{\cal F} }{\partial g^2} 
\ = \ \frac{-H^2 c^{2\ka}}{\ka}\ ,
\label{F''}
\eeq
which is again analytic everywhere. For the third derivative we finally obtain 
\beq
\frac{\partial^3{\cal F} }{\partial g^3}\ = \ \frac{8H^3 c^{3\ka}}{g_2(\ka-2)}
\frac{1}{(c^2-c_c^2)}\ ,
\label{F'''}
\eeq
which has a singularity at $c=c_c$ (and at $\ka=2$ which we can obviously
exclude).  This is precisely where the spectral
density develops an additional zero. We have so far obtained that the
transition is of third order at $c_c$ given by 
eq. (\ref{softcrit}) and thus that the corresponding exponent 
$\ga_{str}$ is negative. 
In order to determine its value 
we still have to find the exponent
$\varepsilon$, with which the denominator vanishes in $g-g_c$. 

Since we cannot directly solve the transcendental equation (\ref{BC}) for $c$
we will expand it around the critical point $c_c$. We make the ansatz
\beq
c^2 \ =\ c_c^2 + \Lambda (g-g_c)^\varepsilon + 
{\cal O}(g-g_c)^{\varepsilon+1} \ ,
\label{cansatz} 
\eeq
where the critical coupling $g_c$ follows from eqs. (\ref{BC}) and
(\ref{softcrit}) to be 
\beq
g_c  \equiv\
\frac{-2c_c^{-\ka}}{H(\ka-2)} \ . 
\label{gac}
\eeq
Inserting the ansatz (\ref{cansatz}) into eq. (\ref{BC}) and expanding we find 
\beq
0\ =\ Hg_c\frac{\ka}{8}(\ka-2)\Lambda^2c_c^{\ka-4}(g-g_c)^{2\varepsilon} +
Hc_c^\ka (g-g_c)\ +\ {\cal O}(g-g_c)^{\varepsilon+1} \ ,
\eeq
which only has a solution for 
\beq
\varepsilon\ = \ \frac12\ .
\eeq
We can also determine the proportionality constant 
\beq
\Lambda^2 \ =\ \frac{4H}{\ka} c_c^{\ka+4} \ ,
\eeq
and we finally obtain 
\beq
c^2 \ =\ c_c^2 \left( 1-\sqrt{4H\ka^{-1}c_c^\ka}\ (g-g_c)^{\frac12} 
+ {\cal O}(g-g_c)^{\frac32}\right)\ .
\eeq
Therefore it follows from eq. (\ref{F'''}) that 
\beq
\frac{\partial^3{\cal F} }{\partial g^3} \sim (g-g_c)^{-\frac12} 
\eeq
and thus $\ga_{str}=-\frac12$ as for the first 
multicritical point of polynomial potentials \cite{Kazakov}.

In summary we have shown that for our prototype example of a Gaussian plus a
real power exactly the same exponent of ${\cal F}$ and number of zeros of
the density at the endpoint $c$ occur, when the potential is tuned to
criticality. The question is if the same is true for higher critical points,
which can be obtained by adding higher powers to the potential. 
From eq. (\ref{rhofreudF2}) it is clear that the spectral density of a 
polynomial plus a Freud type potential 
always has an expansion in powers of $\sqrt{c^2-\la^2}$ and
that thus the accumulation of zeros at $\la=c$ happens the same way as in the
polynomial case. For the free energy however there is no other way to
calculate its exponent than in 
the tedious analysis as presented above, by taking
derivatives and expanding the solution of the normalization condition. 
The method of orthogonal polynomials \cite{Philippe} 
typically used to determine the exponents 
of ${\cal F}$ for arbitrary criticality 
$m$ does not easily apply here since we do not have an explicit
form of the string equation (\ref{string}) at hand for finite $n$. 
But as our example above was arbitrarily chosen (for simplicity) we do not
expect that new critical exponents occur for higher multicriticality compared 
to \cite{Kazakov}.

Since we have calculated the free energy for arbitrary couplings we may also 
ask about the order of the phase transition for the critical points of 
Section \ref{multi} at the origin. The free energy is analytic 
everywhere away from the critical endpoint $c_{c}$ eq. (\ref{c_crit}) which 
differs from the value it takes for criticality at the origin 
(see eq. (\ref{multi1})). However, the fact that ${\cal F}$ is thus analytic 
there does not imply a smooth crossover. In principle we would have to compare 
to the free energy and its derivatives computed on the two-cut side at the 
transition \cite{cmm}.
Our findings in Appendix \ref{App} make this a very delicate task.

\sect{Conclusions}\label{con}   

We have introduced a new class of one-matrix models which contain
nonpolynomial powers in the potential. 
These models do not seem to have an immediate graphical
interpretation in terms of Feynman diagrams. However, they still
share properties of models with polynomial potentials 
such as critical exponents related to Quantum Gravity, while in other aspects
they 
show new features: a spectral density vanishing proportional to
an arbitrary real power at
the origin and a new scaling behavior of the eigenvalues. 

We have first determined the macroscopic large-$N$ spectral density
using saddle-point and orthogonal polynomial techniques, giving
explicit examples such as the simplest nonpolynomial extension of a
Gaussian matrix model.  When taking the microscopic scaling limit at
the origin we could prove that the know massless and massive
universality classes for both the chiral and nonchiral ensembles were
maintained, with the spectral density $\rho(0)$ providing the
universal parameter.

After tuning the coupling constants of the potential to make $\rho(0)$
vanish we found a set of new classes of multicritical matrix models with 
$\rho(\la=0)\sim\la^{\ka-1}$ for arbitrary real $\ka>1$. Our analysis
completes the study of such critical models, 
where previously only multicritical
models with a density vanishing as an even power were known. Similarly to the
findings there for higher critical points additional couplings have to be
introduced and tuned. Our results fall into classes with 
$2m-1<\ka<2m+1$ where $m\in\mathbb{N}$. For each class of criticality a
minimal set of $m$ coupling constants has to be tuned and we explicitly gave
a realization of such potentials. 
The scaling behavior for such multicritical
densities was determined and can be tested on the Lattice in applications 
to the chiral phase transition.

We would like to emphasize the generality of the multicritical models
we have considered. Our results also apply to the case when the
critical exponents of the underlying physical model differ from the
mean field values as we can incorporate arbitrary real positive
exponents. It would be thus very interesting to calculate exactly the
correlation functions for such models as a function of the real
parameter $\ka$. Here, we could only derive an approximate
differential equation for the asymptotic wavefunctions, which enter
the kernel of orthogonal polynomials and thus determine all
correlation functions.  An exact differential equation is currently
known only for even integer values of $\ka$. Another open question
left for further investigation is the order of the corresponding phase
transition in our model. While the continuity of $\ka$ at even integer
values suggests a third order transition the infinite slope of the
first critical density with $1<\ka<2$ may still indicate a lower order
transition.

To this aim 
we have also analyzed the possible critical points of the planar 
free energy in our model.
Nonanalytic parts in its derivatives only occurred at the edge of the
spectrum. For a nonanalyticity to occur at the origin we would have to compare 
with the derivatives of the free energy coming from the two-cut side, which we 
have not been able to determine so far.
When studying criticality by 
the accumulation of extra zeros of the density at the
spectrum edge we recovered the critical exponents known from polynomial 
matrix models. 
A first hope of finding new classes of possibly real valued exponents was not
fulfilled. Such models would represent nonrational conformal field
theories. Obviously more complicated nonpolynomial terms have to be added to
the one-matrix model potential to achieve such goal. This could be thought of 
as a generalization of the known 
two-matrix models which do allow for representations of rational theories, 
with one matrix formally integrated out.

\indent

\noindent
\underline{Acknowledgments}: 
J. Christiansen and K. Splittorff are greatfully acknowledged 
for collaboration at an early stage of this work as well as 
for a critical reading. Furthermore, we wish to thank 
G.M. Cicuta, P.H. Damgaard, P. Di Francesco, H.A. Weidenm\"uller 
and J.F. Wheater for useful discussions
as well as E. Kanzieper and I. Kostov for pointing out some references. 
We also thank R.A. Janik for synchronizing  
with us his publication \cite{Janik} on the same topic.
This work was supported by the European network on ``Discrete Random 
Geometries'' HPRN-CT-1999-00161 
(EUROGRID).


\begin{appendix}
\sect{The two-cut solution}\label{App}

In this Appendix we give the solution for nonpolynomial potentials when the 
support of the eigenvalue density splits into two pieces. Since we restrict 
ourselves to the case of even potentials we will not have to enter the 
subtleties encountered for such multi-cut solutions 
where non-perturbative contributions \cite{BDE} may change the meanfield 
solution \cite{A96}.
Our derivation follows \cite{KFII} (see also \cite{KFrev}) 
and since the formalism has been 
presented in detail in Section \ref{largeN} we shall remain brief here. 
We will calculate only the smoothed macroscopic density 
as well as the free 
energy. As has been noted in \cite{KFII} the higher point correlation 
functions do no longer possess a smooth large-$N$ limit.

The main difference to the single support solution is that the recursion 
coefficients $r_n$ from eq. (\ref{rec}) approach two different rather than 
a single smooth function:
\beq
r_N \ =\ \frac12 \left( c\ -\ (-1)^N d\right)\ ,
\label{rec2}
\eeq
where $c$ and $d$ are the endpoints of the support 
$\sigma=[-c,-d]\cup [d,c]$ with $0<d<c$.
Under the condition eq. (\ref{rec2}) with support $\sigma$
it has been shown in \cite{KFrev} 
that the wave functions will approach the following ``smoothed'' limits:
\beqn
\overline{\psi^{(\al)}_N(\la)^2}  &=& \frac{1}{\pi} 
\frac{|\lambda|}{\sqrt{(c^2-\la^2)(\la^2-d^2)}} 
\ \theta(c^2-\la^2) \theta(\la^2-d^2) 
\ , \label{psi2bar}\\
\overline{\psi^{(\al)}_N(\la)\psi^{(\al)}_{N-1}(\la)} &=& \frac{1}{r_N\pi} 
\frac{\la^2-(-1)^Nc\, d}{\sqrt{(c^2-\la^2})(\la^2-d^2)} \ \theta(c^2-\la^2) 
\theta(\la^2-d^2)
\ .
\label{psibaroe}
\eeqn
Similar to eq. (\ref{psibar}) no restrictions on the potential have been made 
in \cite{KFrev}. 
The auxiliary functions eqs. (\ref{A}) and (\ref{B}) can thus easily be 
obtained:
\beqn
A(\la)\ \equiv\ \frac{1}{N}\overline{A_N(\la)}&=& \frac{r_N}{\pi} 
\int_{d}^{c}\! dt~ \frac{t V'(t) - \la V'(\la)}{t^2-\la^2}
\ \frac{t}{\sqrt{(c^2-t^2)(t^2-d^2)}} ~,
\label{Abar2}\\
B(\la)\ \equiv\ \frac{1}{N}\overline{B_N(\la)}&=& \frac{1}{\pi}  
\int_{d}^{c}\! dt~ \frac{\la V'(t) - t V'(\la)}{t^2-\la^2}
\ \frac{t^2-(-1)^Nc\, d}{\sqrt{(c^2-t^2)(t^2-d^2)}} ~.
\label{Bbar2}
\eeqn
Note the alternating sign which is present in both quantities due to eq. 
(\ref{rec2}). We have used already the symmetry with respect to the origin by 
writing intervals over the positive part of $\sigma$ only. 
Inserting these quantities into eq. (\ref{RN}) which is exact for finite 
$N$ we arrive at the following large-$N$ expression for the 
smoothed, macroscopic density:
\beq
\rho(\la) \ =\  \frac{1}{\pi^2}\ \int_{d}^{c}\! dt~ 
\frac{\la V'(t) - t V'(\la)}{t^2-\la^2}
\sqrt{\frac{(c^2-\la^2)(\la^2-d^2)}{(c^2-t^2)(t^2-d^2)}} \ .
\label{rhoII}
\eeq
The normalization condition for the density can be read off from the 
string equations \cite{KFrev} analogous to eq. (\ref{string}):
\beqn
1 &=& \frac{1}{\pi} \int_d^c dt 
 \frac{t^2V'(t)}{\sqrt{(c^2-t^2)(t^2-d^2)}} \ , \nonumber\\
0 &=& \frac{1}{\pi} \int_d^c dt 
 \frac{V'(t)}{\sqrt{(c^2-t^2)(t^2-d^2)}} \ .
\label{BCII}
\eeqn
To make contact to Section  \ref{multi} we give the two-cut result for the 
same simple example, the potential 
\beq
V(\la) = \frac12 g_2 \la^2 \ +\ g |\la|^{\ka} \ . 
\label{VGF2}
\eeq
Inserting it into the general result eq. (\ref{rhoII}) we see that the 
Gaussian terms cancels and we obtain
\beq
\rho(\la) \ =\  \frac{\kappa g}{\pi^2}\ \int_{d}^{c}\! dt~ 
\frac{t^{\ka-2} - \la^{\ka-2}}{t^2-\la^2} \ \la t\ 
\sqrt{\frac{(c^2-\la^2)(\la^2-d^2)}{(c^2-t^2)(t^2-d^2)}} \ .
\label{rhoGF2} 
\eeq
It is very interesting to compare 
this result at the transition point, 
when the two intervals of $\sigma$ merge, 
with the corresponding density eq. (\ref{rhoGFF}) with a single interval 
support at criticality. 
By sending $d\to 0$ it is easy to see that eq. (\ref{rhoGF2}) 
leads exactly to $\la^2$ times a density of Freud type eq. (\ref{rhofreudF}) 
at shifted $\ka\to\ka-2$ (see also eq. (\ref{preA})). We thus arrive at 
\beq
\rho(\la)|_{d=0}
\ =\ \frac{g \ka}{2\pi}
\tan\left[\frac{\ka\pi}{2}\right] \la^{\ka-1}
\ +\ \frac{g\ka(\ka-2) H c^{\ka-3}}{\pi(\ka-1)(\ka-3)} \la^2\ 
F\left(\frac12,\,
\frac32-\frac{\ka}{2};\,\frac52-\frac{\ka}{2};\,\frac{\la^2}{c^2}\right) ,
\label{rhoIId=0}
\eeq
where we have again chosen non-integer $1<\ka<3$. 
It is easy to see that the normalization conditions eq. (\ref{BCII}) matches  
with the normalization condition eq. (\ref{BCr}) for the single cut 
density together with the multicriticality condition eq. (\ref{multi1}).
Moreover the two densities eqs. (\ref{rhoIId=0}) and 
(\ref{rhoGFF}) agree exactly at multicriticality (\ref{multi1}), 
which follows from the general form  
eq. (\ref{finalrho}) for $m=1$. 
As an immediate consequence on the transition point the two-cut free energy  
following from eq. (\ref{FSP}) is identical to eq. (\ref{Fsol}) together with 
eq. (\ref{multi1}) which has been calculated coming from the one-cut side. 

In order to calculate also derivatives of the free energy in the 
two-cut case we would have to first solve the integral in eq. 
(\ref{rhoGF2}) in order to be able to determine ${\cal F}$ from eq. 
(\ref{FSP}). These integrals are not simple elliptic integrals as encountered 
in \cite{A96}, and the presence of the nonpolynomial terms makes such a 
task fairly difficult.

\end{appendix}

\indent

\newcommand{\NP}[3]{{ Nucl. Phys. }{\bf B#1} (#2) #3}
\newcommand{\PL}[3]{{ Phys. Lett. }{\bf B#1} (#2) #3}
\newcommand{\PR}[3]{{ Phys. Rev. }{\bf #1} (#2) #3}
\newcommand{\PRE}[3]{{ Phys. Rev. }{\bf E#1} (#2) #3}
\newcommand{\IMP}[3]{{ Int. J. Mod. Phys }{\bf #1} (#2) #3}
\newcommand{\MPL}[3]{{ Mod. Phys. Lett. }{\bf #1} (#2) #3}

\end{document}